\documentclass[12pt]{article}
\usepackage{color}
\usepackage{graphicx}

\newcommand{\be}{\begin{equation}}
\newcommand{\ee}{\end{equation}}
\newcommand{\ba}{\begin{eqnarray}}
\newcommand{\ea}{\end{eqnarray}}

\newcommand{\mam}{\mathrm}
\title{Self-Consistent Field study of Polyelectrolyte Brushes}
\author{Hidetsugu Seki\\
Department of Pure and Applied Sciences, University of Tokyo\\
3-8-1, Komaba, Meguro-ku, Tokyo 153-8902, Japan\\
Yasuo Y. Suzuki\\
Faculty of Engineering, Takushoku University\\
815-1, Tatemachi, Hachioji-shi, Tokyo 193-0985, Japan\\
Henri Orland\\
Service de Physique Th\'{e}orique, CEA-Saclay\\
91191 Gif-sur-Yvette Cedex, France}
\date{}
\begin{document}
\maketitle
\begin{abstract}
We formulate a self-consistent field theory for
polyelectrolyte brushes in the presence of counterions.  We
numerically solve the self-consistent field equations and study
the monomer density profile, the distribution of counterions, and the
total charge distribution.  We study the scaling relations for the
brush height and compare them to the prediction of other theories.  We
find a weak dependence of the brush height on the grafting density.
We fit the counterion distribution outside the brush by the
Gouy-Chapman solution for a virtual charged wall.  We calculate the
amount of counterions outside the brush and find that it saturates as the charge of the polyelectrolytes increases.

\end{abstract}

\section{Introduction}

Colloidal particles have various industrial applications.  Their
tendency to flocculate sometimes causes problem in
applications.  Charging colloidal particles is an effective way to
prevent the flocculation by electrostatic repulsions.  However,
such charge effects are easily screened by salt and polar solvents in
the surrounding environment.  Grafting polymers onto the colloid
surface offers another way to prevent the flocculation.  Steric
repulsion induced by the grafted polymers makes colloidal particles repel
each other.

Grafting polyelectrolytes onto the colloid surface offers an even more
efficient way to prevent the flocculation of the colloids in polar
solutions \cite{napper83}.  The stabilization effect arises from both
steric and electrostatic repulsion.  Since the polyelectrolytes can
trap their own counterions in their vicinity, the grafting of
polyelectrolytes generates a layer of locally enhanced counterion
concentration.  The stabilization effect obtained by grafting
polyelectrolytes is thus less sensitive to the salinity of the
surrounding environment than the one obtained by solely charging the
colloids.

Grafted neutral polymers have been studied by Alexander\cite{alexander77}
and de Gennes\cite{degennes80}.  When polymer chains densely grafted onto
an impenetrable planar surface are immersed in a good solvent, the
excluded volume effects strongly stretch the chains and they
form a blob structure.  This particular distribution of
densely grafted polymers is called a ``brush''.  In this neutral
brush regime, they found that the density profile
of the brush is essentially flat and they predicted a linear dependence of
the brush height on the polymerization index and some other scaling 
for the grafting density.

Later, self-consistent field theories were developed for neutral
polymer brushes\cite{edwards65}-\cite{netz03}.  
In these papers, it was shown that polymer
chains are not uniformly stretched, and
the density profile of the brush is not flat but parabolic.

Polyelectrolyte brushes which consist of charged polymer chains
densely grafted onto a surface have been extensively studied by theories 
\cite{miklavic88}-\cite{witte06},
simulations\cite{granfeldt90}-\cite{kumar05} and
experiments\cite{watanabe92}-\cite{ahrens04}.
The free energy of the polyelectrolyte brush consists of three terms in
general: The osmotic pressure of counterions, the Coulomb interactions between
the charges, and the excluded volume interactions of monomers.

Pincus \cite{pincus91} has proposed a scaling theory for the
polyelectrolyte brush assuming that its equilibrium height is
determined by balancing the osmotic pressure of the counterions
with the Gaussian elasticity of the chain.  Scaling laws were presented
for two regimes: (1) the osmotic regime: When the polyelectrolyte brushes
are densely grafted and have a large value of their charge fraction
(strongly charged case), most of the counterions are captured
inside and in the vicinity of the brush due to the strong Coulomb
attraction.  The brush height is roughly equal to the thickness of
the counterion layer and it is independent of the grafting density.
(2) Pincus regime: When the chains are sparsely grafted and have small
values of their charge fraction (weakly charged case), the Coulomb
attraction is relatively weak and the counterions are distributed
far beyond the brush region.  The counterion swelling pressure is reduced
by the fraction captured inside the brush.

The polyelectrolyte brush has been further investigated with the box model
where a flat density distribution is assumed, and a
collapsed regime was introduced \cite{csajka01,csajka01b}.  In this
regime, attractive forces between grafted polyelectrolytes occur
due to the correlations of the monomers and counterions.  The
brush height is determined by balancing the above attractive
forces and the repulsive forces due to the excluded volume effects.

Zhulina \textit{et al.}\cite{zhulina92} developed a self-consistent
analytical model and considered the effect of the excluded volume
interactions in the osmotic regime.  In the osmotic regime, for good
solvents, they obtained a scaling relation which does not agree with that of the
box model.
This work was generalized to poor solvents \cite{ross92,zhulina92}.
Borisov \textit{et al.}  \cite{borisov94} introduced a quasi-neutral
or Alexander regime where the excluded volume effects dominate over
electrostatic effects. The quasi-neutral regime has
the same scaling relation as the neutral brush except for the
dependence on the charge fraction of the chain.  The brush height is
determined by a balance between the excluded volume effects and the
chain elasticity.

Self-consistent field theories have been developed for
polyelectrolytes by several authors.  Borukhov \textit{et al.}
applied a self-consistent field theory to semidilute solutions of
polyelectrolytes and polyampholytes \cite{borukhov98}.  Wang
\textit{et al.} applied it to the interface of phase-separated
polyelectrolyte solutions\cite{wang04,wang05}.  Witte \textit{et al.}
applied it to the mixed polyelectrolyte brushes\cite{witte06}.

The phase diagram for various values of the charge fractions and grafting
densities has been obtained from scaling
theories\cite{borisov94,csajka01} or
simulations\cite{seidel03}.
The Bjerrum length $l_{\mam{B}}$ is a characteristic length scale 
at which the electrostatic interactions between monovalent ions and 
the thermal energy become equal.  When $l_{\mam{B}}^3 > v$, where $v$ is the
excluded volume parameter, the electrostatic interactions are relatively
strong.  This situation is called the 'strong coupling case'.
When $l_{\mam{B}}^3 < v$, the electrostatic interaction is relatively
weak.  This situation is called the 'weak coupling case'.  Scaling theories
\cite{csajka01} and molecular dynamics simulations\cite{csajka01b} both predict
that a collapsed regime appears in the strong coupling case.

In the present work we formulate the self-consistent field theory for
a salt-free polyelectrolyte brush in the presence of counterions
based on a theory developed  by Orland and
Schick\cite{orland96} for the neutral polymer case combined
with a theory developed by Netz and Orland\cite{netz00} for charged
electrolytes.  We apply it to
polyelectrolyte brushes grafted onto an impenetrable planar surface.
The chain
configurations are obtained by solving some modified Edwards equations,
including electrostatic interactions.  The electrostatic
potential is obtained by solving the nonlinear Poisson-Boltzmann
equation.

We solve these self-consistent field equations numerically in the weak
coupling case and obtain the monomer density profile, counterion distribution,
and total charge distribution.
These results are obtained for various charge fractions and we study the
scaling relations resulting from our model.

\section{The Partition Function for Polyelectrolyte Brushes}

We consider $M$ charged polymers of polymerization index $N$ grafted to a planar
surface of area $A$. We assume that the 
grafting surface is impenetrable so that all polymers are
confined to a half-space.  The monomer
positions are denoted by $\{\mathbf{R}_k (s)\}$,
where $s \in [0,N]$ is the curvilinear abscissa
along a chain of $N$ monomers and $k=1,\cdots,M$ is the label of the $M$
chains. In addition, we assume that the $s=0$ extremities of the chains are grafted to the surface, 
at positions $\mathbf{R}_k (0)=( \mathbf{u}_k,z_k(0)=0)$, 
where $z$ denotes the axis perpendicular 
to the grafting plane.
We consider $N_c$ counterions and denote their positions $\mathbf{r}_i $, where
$i=1,\cdots,N_c$. The canonical partition function $Z_{N_c}$
for this system can be written as
\ba
Z_{N_c}&=&\frac{1}{N_c!}\left[\prod_{i=1}^{N_c}\int d\mathbf{r}_\mam{i} \right]
\left[\prod_{k=1}^{M}\int_{\mathbf{R}_k (0)=(\mathbf{u}_k,z_k(0)=0)} 
\mathcal{D} \mathbf{R}_k (s) \right]
\exp \left(-\frac{3}{2a^2}\sum_{k=1}^{M}\int_0^N ds \dot{\mathbf{R}}_k^2(s)
\right) \nonumber \\
& &
\times \exp \left(
-\frac{1}{2 k_B T} \int d\mathbf{r}d\mathbf{r}^\prime
\hat{\rho}_{c}(\mathbf{r})
v_c(\mathbf{r},\mathbf{r}^\prime)
\hat{\rho}_{c}(\mathbf{r}^\prime)
\right) \nonumber \\
& &
\times \exp \left(
-\frac{v}{2}\int d\mathbf{r} \hat{\rho}_\mam{m} ^2(\mathbf{r})
\right),
\label{eq:1}
\ea
where $\int \mathcal{D}g$ is the functional integral over the function
$g(\mathbf{r})$. The first term in the exponent represents the elasticity 
of the Gaussian chain, where 
$\dot{\mathbf{R}}_k (s)$
is the derivative of $\mathbf{R}_k (s)$ with respect to the curvilinear abscissa $s$
and $a$ is the Kuhn length. The grafting of the chains is imposed by the boundary conditions.

The second term represents the electrostatic interaction where
$k_B T$ is the thermal energy,
\be
v_c(\mathbf{r},\mathbf{r}^\prime)=
\frac{1}{4\pi \epsilon}\frac{1}{|\mathbf{r}-\mathbf{r}^\prime|}
\ee
is the Coulomb interaction, $\epsilon$ is the dielectric constant of the 
solvent and $\hat{\rho}_c(\mathbf{r})$ is the local charge density including 
all charges in this system (charged monomers and counterions):
\ba
\hat{\rho}_c(\mathbf{r})&=&
\sum_{k=1}^{M}\int_0^N ds \ q_k (s) \delta(\mathbf{r}-\mathbf{R}_k(s))
+\sum_{i=1}^{N_c}qe \delta (\mathbf{r}-\mathbf{r}_i),
\ea
where $q_k (s)$ is the variable denoting the charge carried by 
the monomer $s$ along the chain $k$. The charge of the counterions is denoted by  $qe$.
Without loss of generality we assume that all polymers are negatively charged.
We use the $smeared$ model for the charge of the chains.
For a polyelectrolyte with a fraction $f$ of its monomers negatively charged,
this model assumes that each monomer carries a uniform fractional charge
$-fe$, where $-e$ is the electron charge. Namely, $q_k(s)=-fe$ for 
any monomer $s$ on any chain $k$.
The condition of electroneutrality is then expressed as
\be
\int d\mathbf{r} \hat{\rho}_c(\mathbf{r})=0,
\ee
which yields $N_c=fMN$.

The third term in equation (\ref{eq:1}) represents
 the excluded volume interactions between monomers, where $v$ 
is the excluded volume parameter and
\be
\hat{\rho}_\mam{m} (\mathbf{r})=
\sum_{k=1}^{M}\int_0^N ds \delta(\mathbf{r}-\mathbf{R}_k(s))
\label{eq:3}
\ee
is the local monomer concentration.

The operator inverse of the Coulomb interaction for a system 
with homogeneous dielectric constant $\epsilon$ is given by
\be
v_c^{-1}(\mathbf{r},\mathbf{r}^\prime)=-\epsilon \nabla^2
\delta(\mathbf{r}-\mathbf{r}^\prime).
\ee
We introduce two fields $\phi_1,\phi_2$ to perform the Gaussian transformations
\ba
& &
\exp \left(
-\frac{1}{2k_B T} \int d\mathbf{r}d\mathbf{r}^\prime
\hat{\rho}_{c}(\mathbf{r})
v_c(\mathbf{r},\mathbf{r}^\prime)
\hat{\rho}_{c}(\mathbf{r}^\prime)
\right) \nonumber \\
& &=\int \mathcal{D}\phi_1 \exp\left\{
-\frac{k_B T}{2}\int d\mathbf{r}d\mathbf{r}^\prime \phi_1(\mathbf{r})
v_c^{-1}(\mathbf{r},\mathbf{r}^\prime)\phi_1({\mathbf{r}^\prime})
+i\int d\mathbf{r} \hat{\rho}_c({\mathbf{r}}) \phi_1({\mathbf{r}})\right\},
\ea
\be
\exp\left[-\frac{v}{2}\int d\mathbf{r}\hat{\rho}_\mam{m} (\mathbf{r})^2 \right]
=\int \mathcal{D}\phi_2 \exp \left[-\frac{1}{2v}
\int d\mathbf{r}\phi_2^2 (\mathbf{r})-i\int d \mathbf{r}\hat{\rho}_\mam{m}(\mathbf{r})
\phi_2 (\mathbf{r})
\right],
\ee
and the partition function reads
\ba
Z_{N_c}&=&\frac{1}{N_c!}\int \mathcal{D}\phi_1 \mathcal{D}\phi_2
\exp 
\left[ 
-\int d\mathbf{r} \frac{\left[\nabla \phi_1 (\mathbf{r}) \right]^2}{8\pi l_B q^2}
-\frac{1}{2v}\int d\mathbf{r}\phi_2^2 (\mathbf{r})
\right]
\left[\int d\mathbf{r}e^{-i\phi_1 (\mathbf{r})} \right]^{N_c} \nonumber \\
& &\prod_{k=1}^M
\left[ 
\int _{ \mathbf{R} (0)=( \mathbf{u}_k,0) }\mathcal{D}\mathbf{R}(s) 
\exp\left[ 
-\int_0^N ds \left\{ 
\frac{3}{2a^2}\dot{\mathbf{R}}^2(s)-i\frac{f}{q}\phi_1 (\mathbf{R}(s))+
i\phi_2 (\mathbf{R}(s))
\right\}
\right]
\right],
\ea
where $l_B=e^2/4\pi \epsilon k_B T$ is
the Bjerrum length.

The partition function is brought into a more tractable form by going to the
grand canonical ensemble
\be
Z_\lambda=\sum_{N_c=0}^\infty \lambda^{N_c} Z_{N_c},
\label{eq:4}
\ee
where $\lambda$ is the fugacity which is related to the particle chemical
potential $\mu$ by $\lambda=e^{\mu/k_B T}$.

The grand canonical partition function can be written as
\be
Z_\lambda=\int \mathcal{D}\phi_1 \mathcal{D}\phi_2
\exp \left[-H\left[\phi_1,\phi_2 \right] \right],
\ee
with the Hamiltonian
\be
\label{ham}
H\left[\phi_1,\phi_2 \right]=\int d\mathbf{r}
\left( 
\frac{[\nabla \phi_1 (\mathbf{r})]^2}{8\pi l_B q^2}
+\frac{1}{2v}\phi_2^2 (\mathbf{r}) -\lambda e^{-i\phi_1(\mathbf{r})}
\right)
-\sum_{k=1}^M \ln Q_k \left[\phi_1,\phi_2 \right],
\ee
where
\be
Q_k \left[\phi_1,\phi_2 \right]=
\int  _{ \mathbf{R} (0)=( \mathbf{u}_k,0) }\mathcal{D}\mathbf{R}(s) \exp \left[
-\int_0^N ds \left\{ 
\frac{3}{2a^2}\dot{\mathbf{R}}^2(s)-i \frac{f}{q}\phi_1(\mathbf{R}(s))
+i\phi_2(\mathbf{R}(s))
\right\}
\right].
\ee
As follows from the definition of the grand canonical partition function, equation (\ref{eq:4}),
the expectation value of the number of counterions is
given by 
\be
\langle N_c \rangle=
\lambda \frac{\partial \ln Z_\lambda}{\partial \lambda}
=\lambda \left\langle 
\int d\mathbf{r}e^{-i\phi_1(\mathbf{r})} \right\rangle.
\label{eq:7}
\ee

In the limit $M \to \infty$, if the grafting points on the surface are random independent variables, we can further simplify eq.(\ref{ham}). Indeed, according to the central limit theorem, we have
\be
\sum_{k=1}^M \ln Q_k \left[\phi_1,\phi_2 \right] \approx_{M \to \infty} M \int d \mathbf{u} P (\mathbf{u})  \ln Q \left[\phi_1,\phi_2 \right] + O(\sqrt{M}),
\ee
where $P (\mathbf{u})$ is the probability distribution of the grafting point $\mathbf{u}$ and  
\be
Q \left[\phi_1,\phi_2 \right]=
\int  _{ \mathbf{R} (0)=( \mathbf{u},0) }\mathcal{D}\mathbf{R}(s) \exp \left[
-\int_0^N ds \left\{ 
\frac{3}{2a^2}\dot{\mathbf{R}}^2(s)-i \frac{f}{q}\phi_1(\mathbf{R}(s))
+i\phi_2(\mathbf{R}(s)).
\right\}
\right].
\ee

In the following, we will assume a uniform grafting density, that is 
\be
P (\mathbf{u})=\frac{1}{A},
\ee
where $A$ is the total grafting area. If we define $\sigma=A/M$ as the average grafting area per chain, we have
\be
\label{part}
Z_\lambda=\int \mathcal{D}\phi_1 \mathcal{D}\phi_2
\exp \left[-
\int d\mathbf{r}
\left( 
\frac{[\nabla \phi_1 (\mathbf{r})]^2}{8\pi l_B q^2}
+\frac{1}{2v}\phi_2^2 (\mathbf{r}) -\lambda e^{-i\phi_1(\mathbf{r})}
\right)
+\frac{1}{\sigma}\int_A d \mathbf{u}  \ln Q \left[\phi_1,\phi_2 \right] 
\right].
\ee
\section{Mean-field Approximation}
In order to evaluate the partition function (\ref{part}), we use the saddle-point method to determine the fields $\phi_1$ and $\phi_2$ which most contribute to the functional integral.
We assume that these two fields possess translational
symmetry in the $(x,y)$ directions, parallel to the grafted plane (remember that $z$ is the direction normal to the plane). 
The Hamiltonian reads
\be
H\left[\phi_1,\phi_2 \right]=A \int dz
\left( 
\frac{1}{8\pi l_B q^2}\left(\frac{d\phi_1(z)}{dz} \right)^2
+\frac{1}{2v}\phi_2^2 (z) -\lambda e^{-i\phi_1(z)}
\right)
-M \ln Q \left[\phi_1,\phi_2 \right],
\ee
where
\be
Q \left[\phi_1,\phi_2 \right]=
\int_{z(0)=0} \mathcal{D}z(s) \exp \left[
-\int_0^N ds \left\{ 
\frac{3}{2a^2}\dot{z}^2(s)-i \frac{f}{q}\phi_1(z(s))
+i\phi_2(z(s))
\right\}
\right].
\ee

The saddle-point method provides the mean-field approximation to the system.
The mean-field $\Phi_1(z),\Phi_2(z)$ are given by the stationarity conditions
\be
\left. \frac{\delta H[\phi_1,\phi_2]}{\delta{\phi_1(z)}} \right|_{\phi_1=-i\Phi_1}=0,
\ee
\be
\left. \frac{\delta H[\phi_1,\phi_2]}{\delta{\phi_2(z)}} \right|_{\phi_2=-i\Phi_2}=0.
\ee
The mean-field approximation to the Hamiltonian $H_{\mathrm{mf}}$ is simply
\be
H_{\mathrm{mf}}\equiv H[i\Phi_1,i\Phi_2].
\ee
Note that $\Phi_1(z)$ is identified as the electrostatic potential.

The mean-field equations are thus
\be
\frac{1}{4\pi l_B q^2}\frac{d^2 \Phi_1(z)}{dz^2}
+\lambda e^{-\Phi_1(z)}-\frac{1}{\sigma}
\frac{\delta \ln Q}{\delta \Phi_1(z)}=0,
\label{eq:mf1}
\ee
\be
\Phi_2(z)=-\frac{v}{\sigma}\frac{\delta \ln Q}{\delta \Phi_2(z)}
\equiv \frac{v}{\sigma}\rho_\mam{m}(z),
\label{eq:6}
\ee
where $\sigma=A/M$ is the average area per chain and $\rho_\mam{m}(z)$ is 
the local density of monomers.

The density of monomers can be written as
\be
\rho_\mam{m} (z)=\frac{1}{Q}\int \mathcal{D}z(s) 
\int_0^N ds \delta(z-z(s))
\exp \left[
-\int_0^N ds \left\{ 
\frac{3}{2a^2}\dot{z}^2(s)-\frac{f}{q}\Phi_1(z(s))
+\frac{v}{\sigma}\rho_\mam{m}(z(s))
\right\}
\right].
\ee
The density is normalized so that
\be
\int_0^\infty \rho_\mam{m}(z)dz=N.
\ee

We take the grafting plane to be at $z=0$ and 
introduce the monomer probability amplitudes
\be
\psi(z,s)=\langle z| e^{-sH}|0 \rangle ,
\ee
\be
\psi^\dagger(z,s)=\int_0^\infty
dz^\prime \langle z^\prime | e^{-sH}| z \rangle
\ee
where we use the standard quantum-mechanical notations
\be
\langle z^\prime | e^{-sH}| z \rangle ,
=
\int_{z(0)=z}^{z(s)=z^\prime}
\mathcal{D}z(s) \exp \left[
-\int_0^s ds^\prime \left\{ 
\frac{3}{2a^2}\dot{z}^2(s^\prime)-\frac{f}{q}\Phi_1(z(s^\prime))
+\frac{v}{\sigma}\rho_{\mam{m}}(z(s^\prime))
\right\}
\right].
\ee
Note that as before in the paper, the curvilinear abscissa  $s$ is defined so that its values at the
grafted and free ends are $0$ and $N$ respectively.

The probability amplitudes $\psi(z,s),\psi^\dagger (z,s)$ satisfy the
modified diffusion equations
\be
\left[
\frac{\partial}{\partial s}-\frac{a^2}{6}\frac{\partial^2}{\partial z^2}
-\frac{f}{q}\Phi_1(z)+\frac{v}{\sigma}\rho_\mam{m}(z)
\right] \psi(z,s)=0 ,
\label{eq:8}
\ee
\be
\left[
\frac{\partial}{\partial s}-\frac{a^2}{6}\frac{\partial^2}{\partial z^2}
-\frac{f}{q}\Phi_1(z)+\frac{v}{\sigma}\rho_\mam{m}(z)
\right] \psi^\dagger(z,s)=0.
\label{eq:9}
\ee
with the initial conditions $\psi(z,0)=\delta(s)$ and $\psi^{\dagger} (z,0)=1.$
Because the grafting plane is impenetrable,$\psi(z,s),\psi^\dagger (z,s)$ must satisfy
the boundary conditions $\psi(0,s)=0,\psi^\dagger(0,s)=0$.

The density of monomers
is given by
\be
\rho_\mam{m}(z)=
\frac{\int_0^N ds \psi(z,s)\psi^\dagger (z,N-s)}
{\int_0^\infty dz^\prime \psi(z^\prime,N)},
\label{eq:11}
\ee
and equation (\ref{eq:mf1}) can be written as
\be
\frac{1}{4\pi l_B q^2}\frac{d^2 \Phi_1(z)}{dz^2}
+\lambda e^{-\Phi_1(z)}-\frac{f}{q}\frac{1}{\sigma}\rho_\mam{m}(z)=0.
\label{eq:mf11}
\ee
This is just the Poisson-Boltzmann equation for all the charges in the system
 (polymers and counterions). In the following,  we will consider only monovalent counterions, that is $q=1$.

The counterion density per unit area of the grafting plane is given by
\be
\rho_{\mathrm{ci}}(z)=\lambda e^{-\Phi_1(z)}.
\ee
From equation (\ref{eq:mf11}) one obtains
$
\int_0^\infty \rho_{\mathrm{ci}}(z)dz=\langle N \rangle/A.
$
The condition of electroneutrality is given by
\be
\int_0^\infty \rho_{\mathrm{ci}}(z)dz =\frac{fN}{\sigma}.
\label{eq:12}
\ee

Equations (\ref{eq:8}) to (\ref{eq:12}) are the self-consistent field equations describing 
the polyelectrolyte brush in the presence of counterions.
We assume that the grafting plane is electrically neutral. The electroneutrality of the
system is enforced by setting the field strength to zero at the boundaries
\be
\left. \frac{d \Phi_1 (z)}{dz} \right|_{z=0}
=\left. \frac{d \Phi_1(z)}{dz}\right|_{z=\infty}=0.
\label{eq:13}
\ee
The total charge density $\rho_{\mam{c}}(z)$ per unit area of the grafting plane 
is given by
\be
\rho_{\mam{c}}(z)=-\frac{f}{\sigma}\rho_{\mam{m}}(z)+\rho_{\mam{ci}}(z).
\label{eq:15}
\ee
The total charge distribution $\rho_{\mam{c}}(z)$ satisfies the condition of 
the electroneutrality
\be
\int_0^\infty \rho_{\mam{c}}(z) dz=0.
\label{eq:14}
\ee

\section{Numerical Solutions of the Self-Consistent Field Equations and Discussions}

In this section we present numerical solutions of the self-consistent field
equations as a function of the values of the polymerization index $N$, 
grafting area per chain $\sigma$, excluded volume parameter $v$,
charge fraction $f$, and Bjerrum length $l_{\mam{B}}$.
In this paper, we vary $f$ from $0$ to $1.0$,
while we fix the values of parameters $N=30, v=1.0
a^3, l_{\mam{B}}=0.1a, \sigma=10a^2$, otherwise stated.
This corresponds to the weak coupling case $ l_{\mam{B}}^3 < v$ 
where the Coulomb interaction is relatively weak.  
To investigate the grafting density dependence, we vary
$\sigma$ from $5 a^2$ to $20 a^2$.  To investigate the polymerization index
dependence, we vary $N$ from $30$ to $70$.

\subsection{Structure of the Brush}

\begin{figure}[htbp]
 \begin{center}
  \rotatebox{-90}{
    \includegraphics[width=8cm]{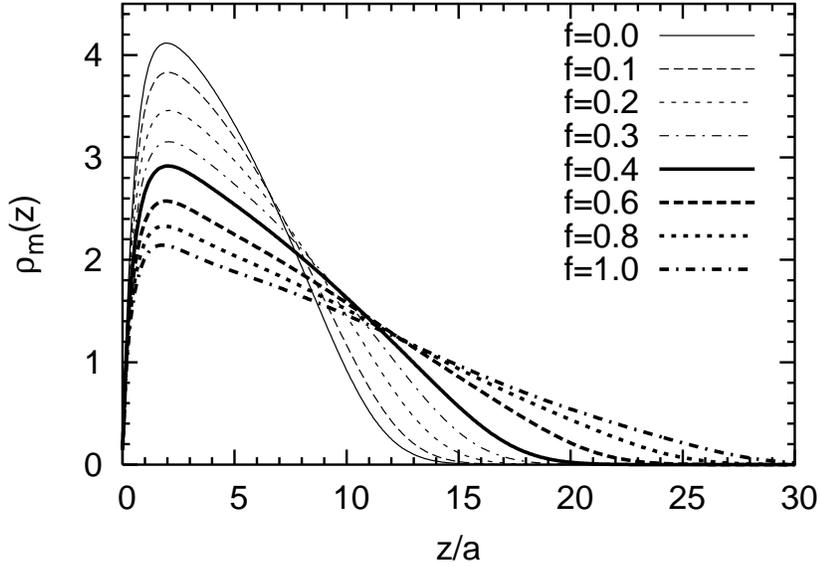}}
  \caption{Monomer density profile $\rho_\mathrm{m}(z)$ for polymerization index $N=30$, 
           grafting area per chain $\sigma=10 a^2$,  excluded volume parameter $v=1.0 a^3$, 
   Bjerrum length $l_{\mam{B}}=0.1 a$. Profiles for the charged fraction 
           $f$=$0(\mam{neutral}), 0.1, 0.2, 0.3, 0.4, 0.6, 0.8, 1.0$ are shown.}
 \label{figure1}
 \end{center}
\end{figure}

In Figure \ref{figure1} we show the monomer density profiles
$\rho_{\mam{m}}(z)$ for the above fixed values and various values of $f$
including the monomer density profile of the neutral brush ($f=0$). 
The brush height increases with increasing $f$.

The impenetrability of the grafting plane induces a depletion area near
the grafting surface in the monomer density profiles and pushes the maximum
of the density away from the grafting surface.
The depletion area is also observed for the neutral brush\cite{scheutjens79}.
The density maximum gets closer to the grafting plane with increasing $f$.  
This is due to the strong Coulomb interaction between the monomers and the
counterions for large $f$.  The monomer density decreases
exponentially when the distance from the grafting plane $z$ becomes
large. 

Taniguchi \textit{et al.}\cite{taniguchi03} obtained similar results
but the behavior of the monomer density near the grafting plane is
somewhat different from our results.

We measure the brush height $h_{\mam{m}}$ and the thickness of the counterion layers
$ l_{\mam{ci}} $ as the first moments of each density profiles:
\be
h_{\mam{m}}=\frac{\int_0^\infty z \rho_{\mathrm{m}}(z)dz}
{\int_0^\infty  \rho_{\mathrm{m}}(z)dz},
\ee
\be
l_{\mam{ci}}=\frac{\int_0^\infty z \rho_{\mathrm{ci}}(z)dz}
{\int_0^\infty \rho_{\mathrm{ci}}(z)dz}.
\ee
We show the values of $2h_{\mam{m}}$ and $2l_{\mam{ci}}$ for each values of $f$
in Table \ref{table1}.  We multiply by a factor $2$ because the height
for the uniform density distribution corresponds to twice of the first moment.

\begin{figure}[htbp]
 \begin{center}
 \begin{tabular}{c}
   \rotatebox{-90}{
     \includegraphics[width=8cm]{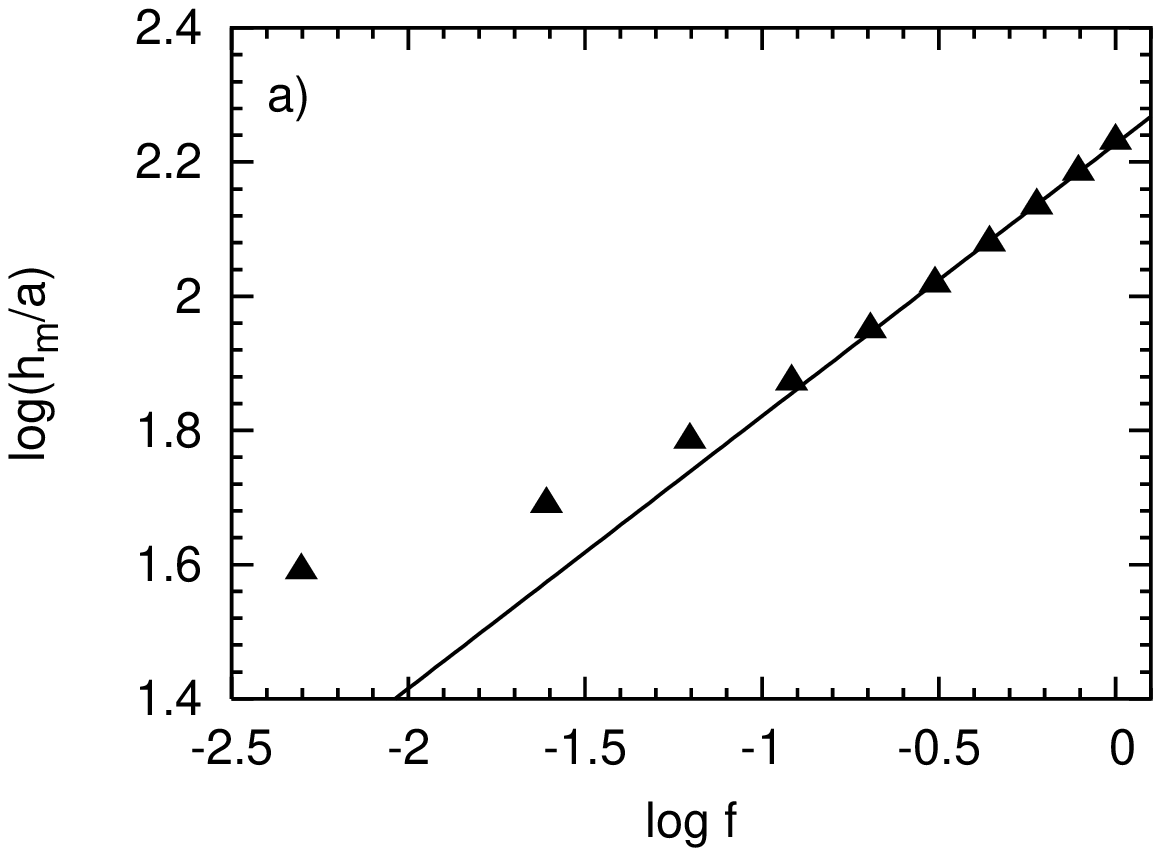}}\\
   \rotatebox{-90}{
     \includegraphics[width=8cm]{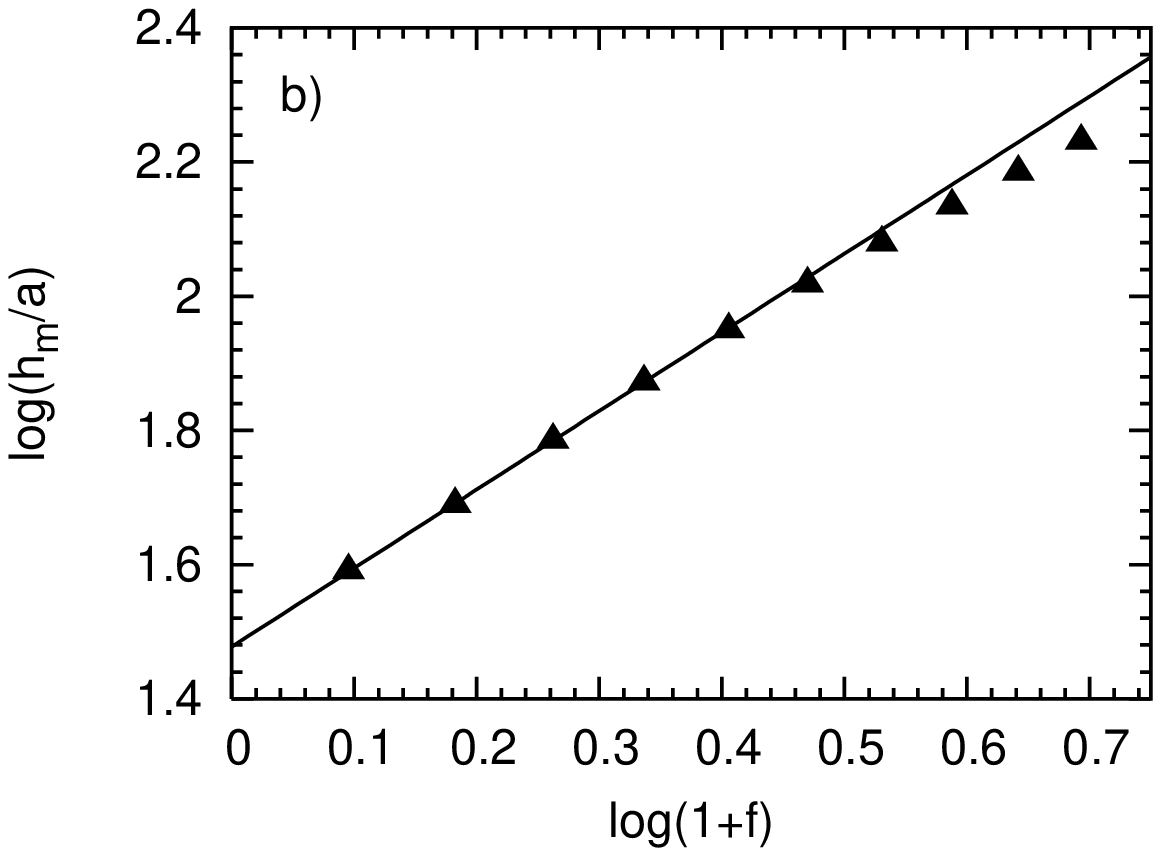}}
  \end{tabular}
    \caption{(a)
 The brush height $h_\mam{m}$ as a function of the charge fraction $f$.
  The parameters are the same as in Figure \ref{figure1}.
  We plot $f=0.1, 0.2, \cdots, 0.9, 1.0$ from left to right.  solid line shows a power
  law fit.   
    (b)The brush height $h_\mam{m}$ as a function of $1+f$.
  The parameters are the same as in Figure \ref{figure1}.
  We plot $f=0.1, 0.2, \cdots, 0.9, 1.0$ from left to right.The solid line shows a power
  law fit.
        }
    \label{figure2}
  \end{center}
 \end{figure}

In Figure \ref{figure2}a we plot $h_{\mam{m}}$ and $f$ on a log-log scale.
For large values of $f \ge 0.5$, the brush height behaves as
$h_{\mam{m}} \sim f^\alpha$, with exponent $\alpha = 0.41$.
In Figure \ref{figure2}b we plot $h_{\mam{m}}$ and $1+f$ on a log-log scale.
Figure \ref{figure2}b shows the scaling relation 
$ h_{\mam{m}} \sim (1+f)^\beta $ with exponent $\beta = 1.17$ for small values
of $f \le 0.5$.  

\begin{figure}[htbp]
 \begin{center}
   \rotatebox{-90}{
     \includegraphics[width=6cm]{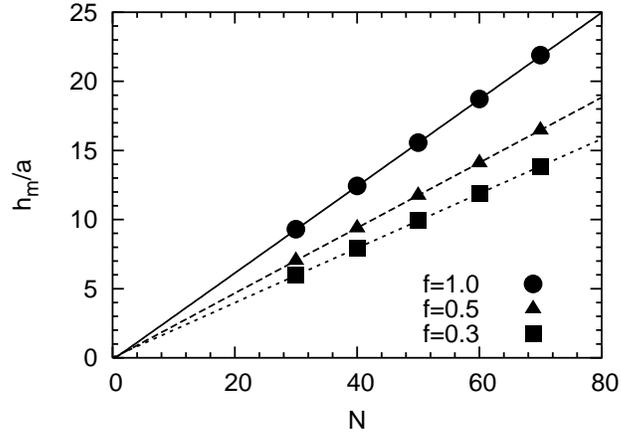}}\\
    \caption{
 Brush height $h_\mam{m}$ as a function of the polymerization index $N$.
 $f=1.0$(circles), $f=0.5$(triangles), $f=0.3$(squares) are shown.
 The other parameters are the same as in Figure \ref{figure1}.
        }
    \label{figure3}
  \end{center}
 \end{figure}

\begin{figure}[htbp]
 \begin{center}
  \rotatebox{-90}{
    \includegraphics[width=6cm]{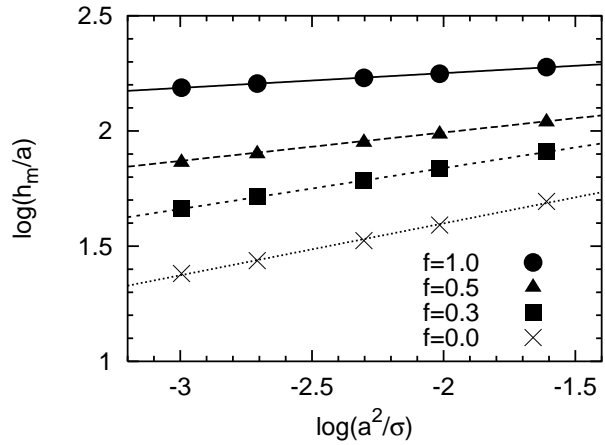}}\\
  \caption{Brush height $h_\mam{m}$ as a function of the grafting density $1/\sigma$.
           $f=1.0$(circles), $f=0.5$(triangles), $f=0.3$(squares), $f=0.0$(cross).
           The other parameters are the same as in Figure \ref{figure1}.}
   \label{figure4}
 \end{center}
\end{figure}

We investigate the dependence of the brush height $h_{\mam{m}}$ on the other parameters
$N$ and $\sigma$.  
Figure \ref{figure3} shows the scaling relation $h_{\mam{m}} \sim N^\gamma$
with exponent $\gamma = 1.0$ for $f=0.3,0.5,1.0$ respectively.
Figure \ref{figure4} shows the scaling relation $ h_{\mam{m}} \sim (1/\sigma)^\delta $ with exponent $\delta = 0.06, 0.13, 0.18, 0.23$ for $f=1.0, 0.5, 0.3, 0.0$ respectively.

We briefly review the scaling relation predicted by the scaling theory and other SCF theory.
For the strongly charged case, Pincus\cite{pincus91} predicted that the brush height $h$ 
behaves as
\ba
&& h \sim N a f^{1/2} , \label{eq:r1} \\ 
&& \alpha=1/2, \gamma=1, \delta=0 , \nonumber
\ea 
in the osmotic regime.

Zhulina \textit{et al.}  \cite{zhulina92} found that the scaling relation in
the osmotic regime changes due to excluded volume effects to
\ba 
&& h \sim N a f^{2/5} , \label{eq:r2} \\ 
&& \alpha=2/5, \gamma=1, \delta=0 . \nonumber
\ea
The scaling relations (\ref{eq:r1}), (\ref{eq:r2}) indicate
that the brush height $h$ is independent of the grafting density $1/\sigma$.

Recent study for the osmotic regime shows that the brush height $h$ slightly 
increases with increasing $1/\sigma$\cite{naji03,seidel03,romet04,ahrens04}.
The term `nonlinear osmotic regime' is used for this weak dependence on $1/\sigma$. ($\delta=1/5$ for weak coupling and $f=1$) 

Pincus\cite{pincus91} also obtained the brush height 
in the Pincus regime
where counterions are distributed beyond the brush  
behaves as
\ba 
&& h \sim N^3 a^2 f^2 /\sigma ,   \\ 
&& \alpha=2, \gamma=3, \delta=1 .  \nonumber
\label{eq:r3}
\ea

The quasi-neutral regime\cite{alexander77,degennes80,borisov94,csajka01} 
appears 
both for the strongly and weakly charged case.  The
brush height is determined by balancing the repulsive force due to the
excluded volume effect and the attractive force due to the chain
elasticity.  The scaling relation is identical to the neutral brush
case except for the dependence on $f$ for the strongly charged case.
The scaling relation for the strongly charged case is 
\ba 
&& h \sim (1+f)^{2/3} N (a^2 v/\sigma)^{1/3} , \\ 
&& \beta=2/3, \gamma=1, \delta=1/3 .  \nonumber
\label{eq:r4}
\ea
For the weakly charged case, the scaling relation is exactly the same
as that for the neutral brush:
\ba
&& h \sim N (a^2 v/\sigma)^{1/3} ,  \\ 
&& \alpha=0, \gamma=1, \delta=1/3 .  \nonumber
\label{eq:r5}
\ea

We compare our numerical results with the other theoretical
predictions.  First we investigate the results for $f \ge 0.5$.  Our
numerical results $\alpha = 0.41$ is in good agreement with
$\alpha=2/5$ of (\ref{eq:r2}) for the osmotic regime in the good
solvent.  Our numerical results $\gamma=1.0$ for $f=0.5,1.0$ is
consistent with (\ref{eq:r2}).  The brush height scales linearly with
$N$.  Our results show that the brush height $h_{\mam{m}}$ depends
weakly on the grafting density $1/\sigma$ with an exponent
$\delta=0.06,0.13$ for $f=1.0,0.5$ respectively.  The scaling relation
(\ref{eq:r2}) predicts $\delta=0$.  These comparisons show that our
numerical results for $f \ge 0.5$ corresponds to the osmotic regime.
The slight increase of the brush height for increasing the grafting
density is consistent with a recent study of the nonlinear osmotic
regime.\cite{naji03,seidel03,romet04,ahrens04}.

Next we investigate our results for $f \le 0.5$. 
Our results of $\beta=1.17$
does not agree with the value $\beta=2/3$ of the scaling relation (\ref{eq:r4}) for
the quasi-neutral regime.  Our results of $\gamma=1.0$ for $f=0.3$ 
is consistent with (\ref{eq:r4}).  Our results of $\delta=0.18,0.23$
for $f=0.3,0.0$ differ from the exponent $\delta=1/3$ of (\ref{eq:r4}).

The neutral brush is characterized by a single stretching parameter\cite{netz98,netz03}
\be
b=N \left(\frac{3 v^2}{2 a^2 \sigma^2}\right)^{1/3}.
\ee
For large value of  $b \ge 10$, the monomer density profile 
is well approximated by the semi classical parabola density profile\cite{milner88}
and the brush height satisfies the scaling relation (\ref{eq:r5}).
However the deviations from the parabolic profile become important when the values of 
$N,1/\sigma,v$ decrease and the stretching parameter becomes $b \le 10$.
Our values of the parameters 
$ N=30,v=1.0 a^3,5 a^2 \le \sigma \le 20 a^2 $ correspond to 
$ 11.7 \ge b \ge 4.7 $.
Therefore, the disagreement between the calculated value of the exponent $\delta$ (0.06, 0.13, 0.18, 0.23) 
and the value of the scaling prediction (\ref{eq:r5}) ($\delta=1/3$) is not surprising.

When $f$ decreases, the value of the exponent $\delta$ becomes close 
to the value for the neutral case.  Roughly speaking 
our results for small value of $f \le 0.3$ correspond to the quasi-neutral regime of the 
phase diagram\cite{borisov94, csajka01, seidel03}.

Our value for the exponent $ \beta=1.17 $ does not agree with the value $\beta=2/3$
of the scaling relation (\ref{eq:r4}). 
When (\ref{eq:r4}) was derived, only excluded volume interactions were considered and 
the osmotic pressure of counterions was neglected.
A more realistic value of the exponent $ \beta $ should be derived when the
excluded volume interaction and the osmotic pressure of counterions are
equally treated.

Our numerical results show a
crossover phenomena
between the osmotic regime and the quasi-neutral regime.
Scaling theories assumes that one of the two repulsive effects is dominant and balance it to the elasticity of the Gaussian chain.
But in many regimes the two effects are comparable; The behavior of
the brush does not change abruptly and a smooth transition between the two regime is expected.
Our numerical results for $0.4 \le f \le 0.5$ correspond to this crossover region.

\subsection{Distribution of Counterions and Charges}

\begin{figure}[htbp]
 \begin{center}
   \rotatebox{-90}{
     \includegraphics[width=8cm]{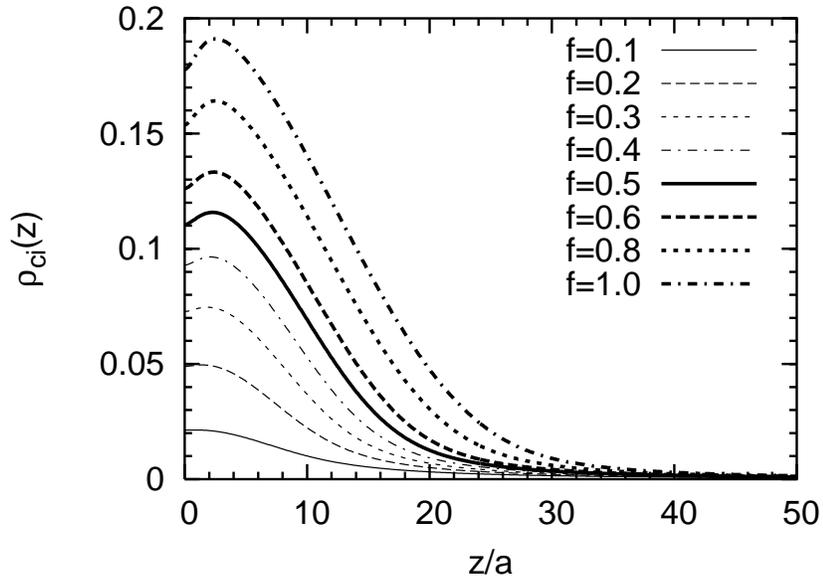}}\\
    \caption{The counterion distribution $\rho_{\mam{ci}}(z)$ for
      $f=0.1, 0.2, 0.3, 0.4, 0.5, 0.6, 0.8, 1.0$. The other parameters are the same as in Figure \ref{figure1}.
   }
    \label{figure5}
  \end{center}
 \end{figure}

Next we consider the distribution of counterions and charges.
We first plot the distribution of counterions $\rho_{\mam{ci}}(z)$
for various values of $f$ in Figure \ref{figure5}.
We observe a depletion area of $\rho_{\mam{ci}}(z)$ near the grafting surface.  
This depletion effect is clear for $f \ge 0.3$. 

At large distance from the wall, the monomer density decreases faster than the counterion density. Therefore there are no more monomers in the solution, and the counterions see a virtual charged wall with an effective charge density $\sigma_{\rm eff}$ at position $r_{\rm eff}$. 

We study the thickness of the counterion layer.
In Table \ref{table1}, we show the values of $2 l_{\mam{ci}}$ for each values 
of $f$.

We fit the counterion distribution at large distance 
by the GC profile
\be
f(z)=\frac{1}{2 \pi l_{\mam B} (z-L_{\mam{f}})^2},
\label{eq:r6}
\ee
where $L_{\rm f}$ is a fit parameter.

For $z \gg L_{\mam{f}}$, eq.(\ref{eq:r6}) is equivalent to the 
distribution of counterions in a system which consists of a virtual charged plane located at $r_{\rm eff}$ with charge density $-\sigma_{\mam{eff}}e$ in the presence of
monovalent counterions. 
The parameter $r_{\rm eff}$ is set to be the minimum value which satisfies the condition
\be
\int_0^{r_{\rm eff}} \rho_c(z) dz = - \int_{r_{\rm eff}}^{\infty} f(z)dz,
\ee
which expresses the fact that the total charge to the left of $r_{\rm eff}$ is equal to that of the counterions on the right of $r_{\rm eff}$.
$\sigma_{\rm eff}$ is determined by 
\be
\sigma_{\rm eff}= \frac{1}{2 \pi l_{\rm B} (r_{\rm eff} - H_{\rm m})}.
\ee

The values of $r_{\rm eff}, \sigma_{\mam{eff}}$ are presented in
Table \ref{table1}.  The effective charge density $\sigma_{\mam{eff}}e $ is almost
equal to the integrated amount of charges of counterions for 
$z \ge r_{\rm eff}$.  The total amount of charges of counterions
is $\sigma_{\mam{s}}\equiv f N/\sigma$.
From $\sigma_{\mam{eff}}$ we define the effective Gouy-Chapman (GC) length as
$l_{\mam{eff}}=1/2\pi l_{\mam{B}}\sigma_{\mam{eff}}$.
This effective GC length  $l_{\mam{eff}}$ provides a definition of the
thickness of the counterion layer outside the virtual wall.
We may thus take $ L_{\mam{ci}} \equiv r_{\rm eff}+l_{\mam{eff}} $ as an alternative definition of the  
thickness of the counterion layer.

\begin{figure}[htbp]
 \begin{center}
 \begin{tabular}{c}
   \rotatebox{-90}{
     \includegraphics[width=6cm]{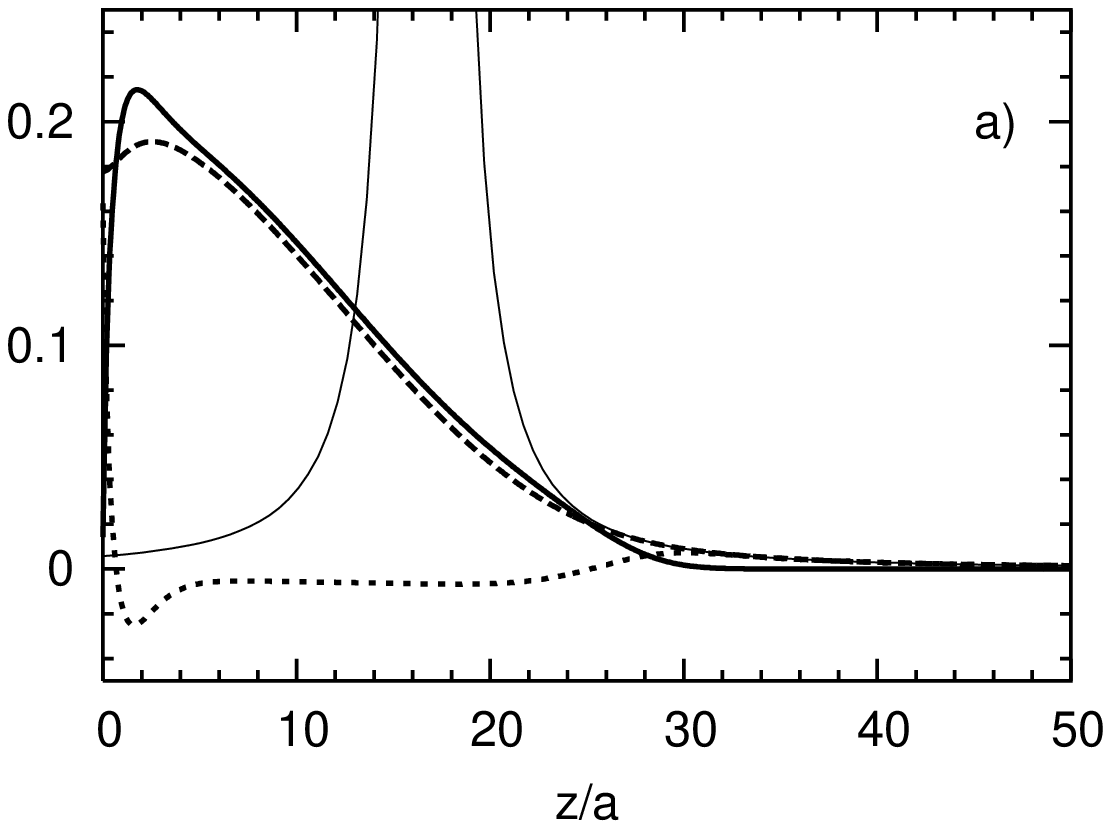}}\\
   \rotatebox{-90}{
     \includegraphics[width=6cm]{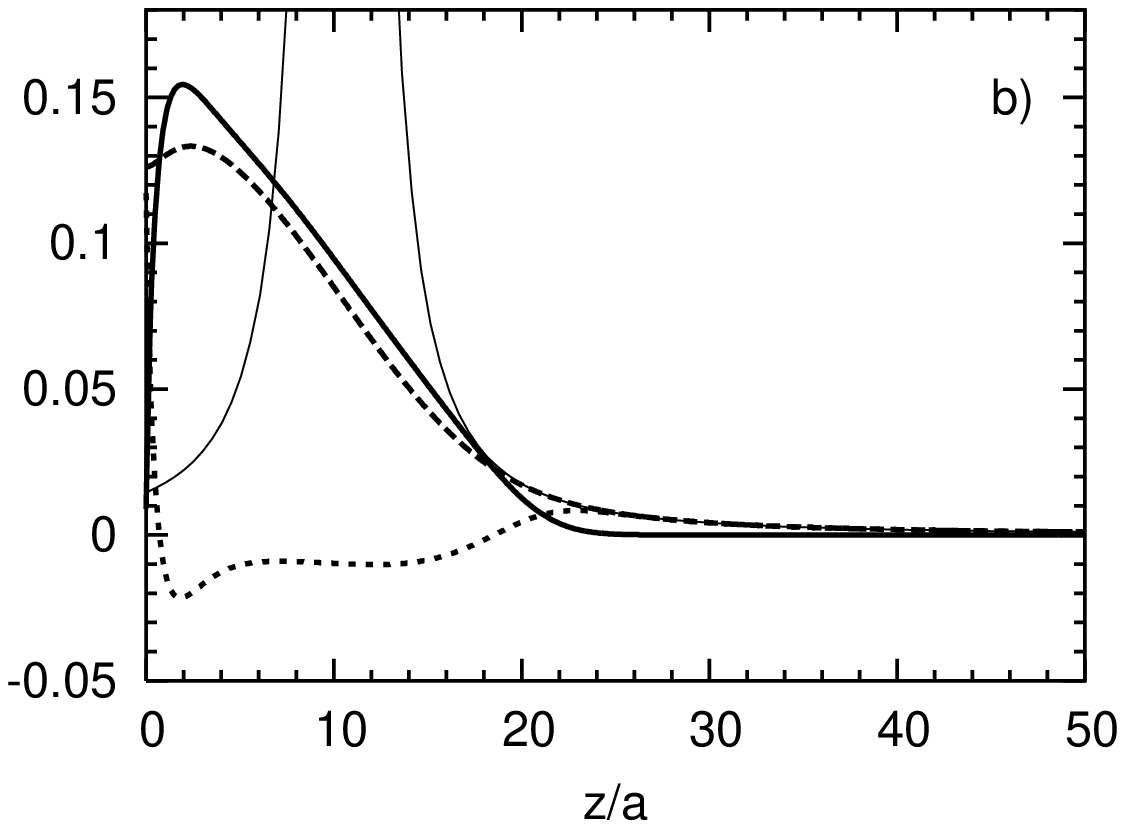}}\\
   \rotatebox{-90}{
     \includegraphics[width=6cm]{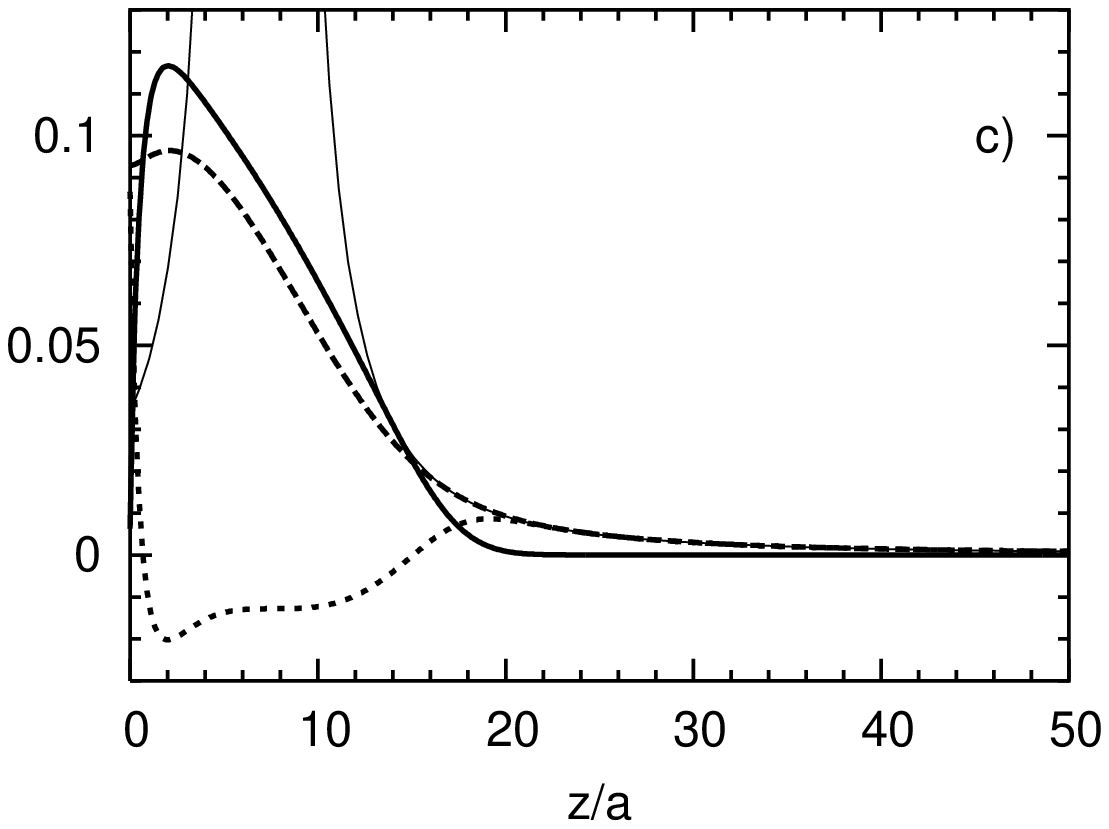}}\\  
  \end{tabular}
    \caption{The thick solid lines are absolute values of the negative charge distributions 
             $fN/\sigma \rho_\mam{m}(z)$,  the
             thick dashed lines represent the positive charge distribution $\rho_{\mam{ci}}(z)$,  the
             thick dotted lines are the total charge distributions $\rho_{\mam{c}}(z)$ ,  and 
             solid lines are fitting curves for the Gouy-Chapman solution.
             The parameters are the same in Figure \ref{figure1}.
             (a) f=1.0, (b) f=0.6, (c)f=0.4.}
    \label{figure7}
  \end{center}
 \end{figure}

\begin{figure}[htbp]
 \begin{center}
 \begin{tabular}{c}
   \rotatebox{-90}{
     \includegraphics[width=6cm]{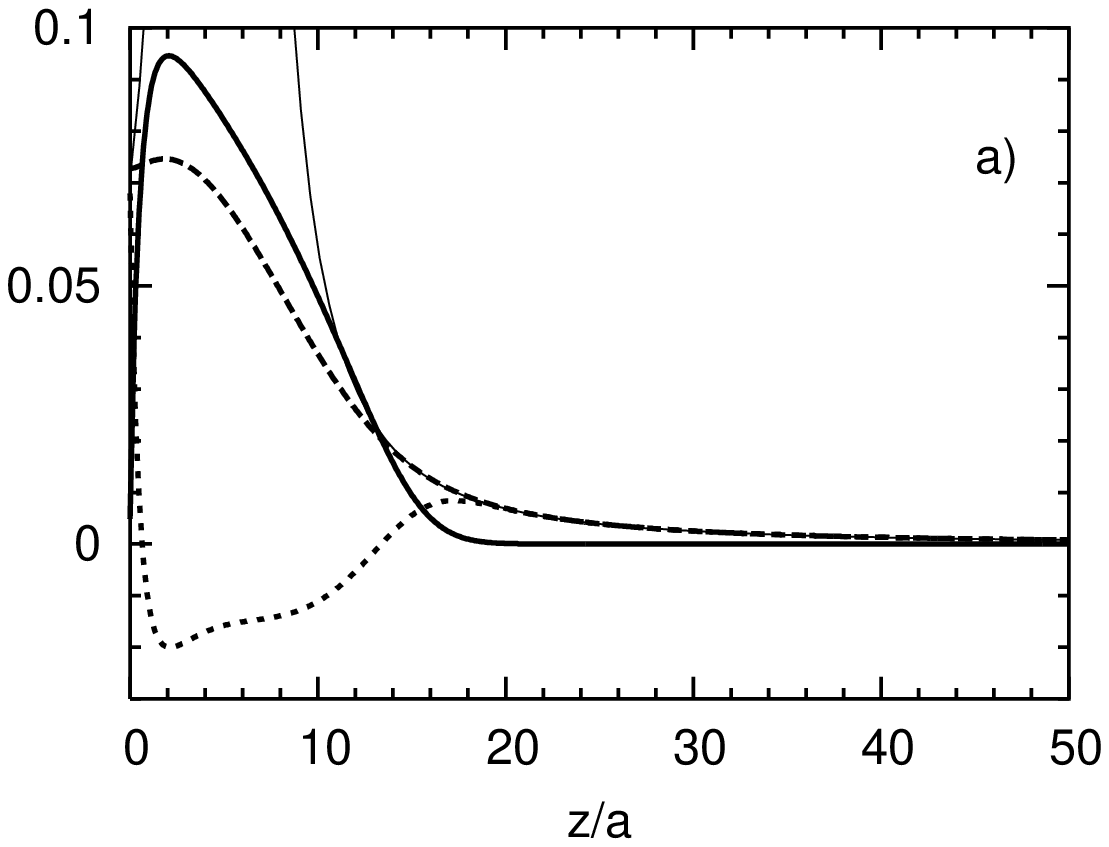}}\\
   \rotatebox{-90}{
     \includegraphics[width=6cm]{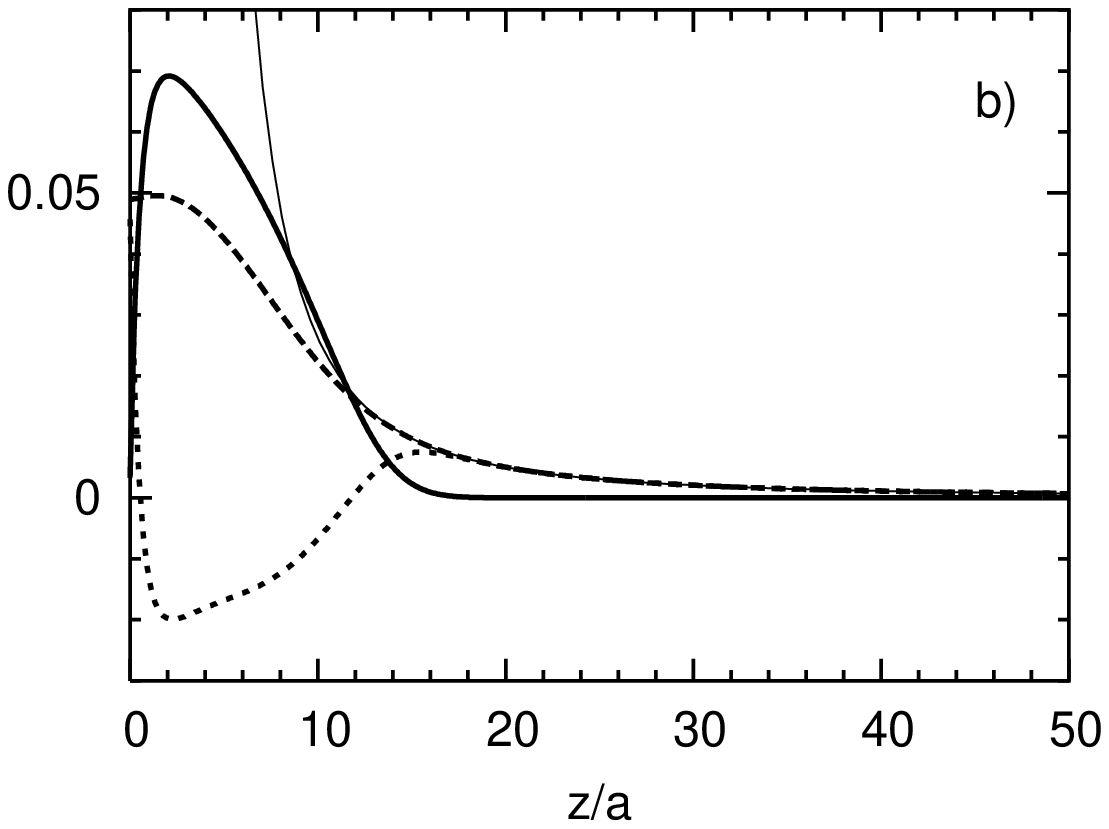}}\\
   \rotatebox{-90}{
     \includegraphics[width=6cm]{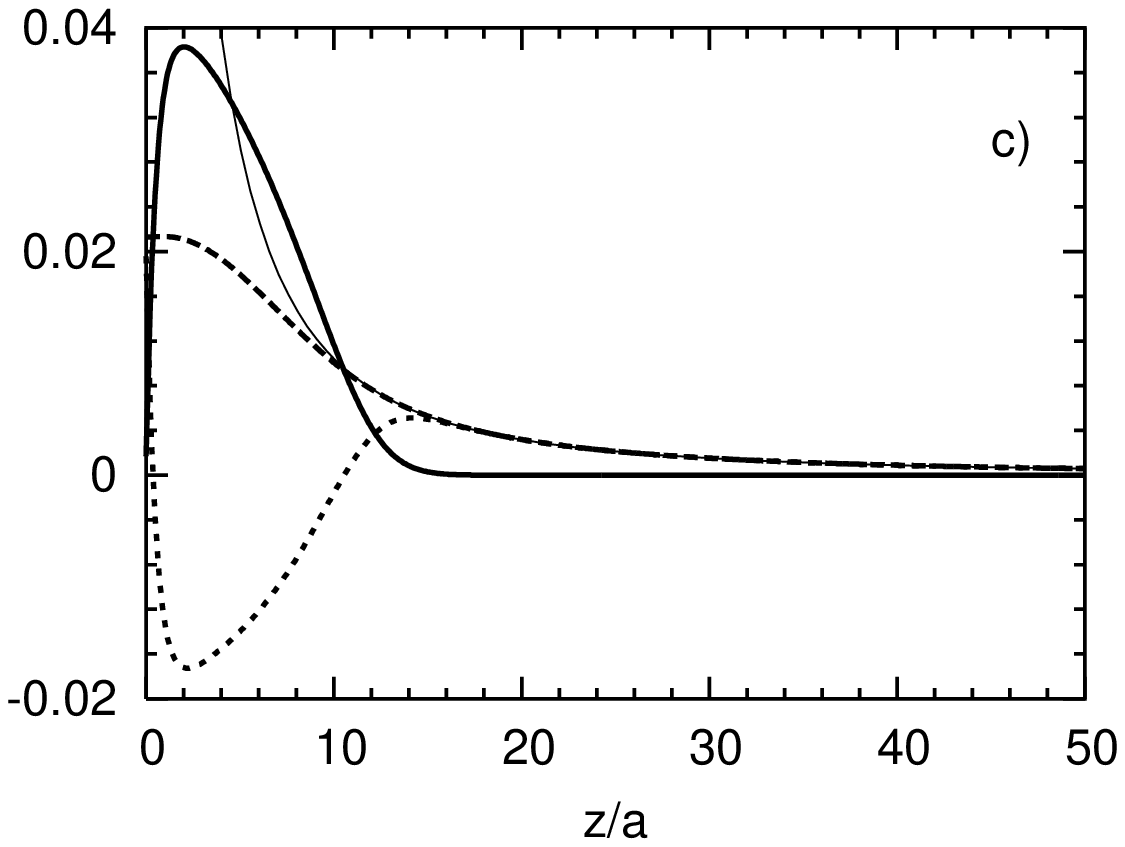}}\\  
  \end{tabular}
    \caption{The thick solid lines are absolute values of the negative charge distributions 
             $fN/\sigma \rho_\mam{m}(z)$, 
             thick dashed lines are the positive charge distributions $\rho_{\mam{ci}}(z)$,  the
             thick dotted lines are the total charge distributions $\rho_{\mam{c}}(z)$ ,  and the
             solid lines are fitting curves for the Gouy-Chapman solution.
             The parameters are the same in Figure \ref{figure1}.
             (a) f=0.3, (b) f=0.2, (c)f=0.1.}
    \label{figure8}
  \end{center}
 \end{figure}

\begin{figure}[htbp]
 \begin{center}
   \rotatebox{-90}{
     \includegraphics[width=8cm]{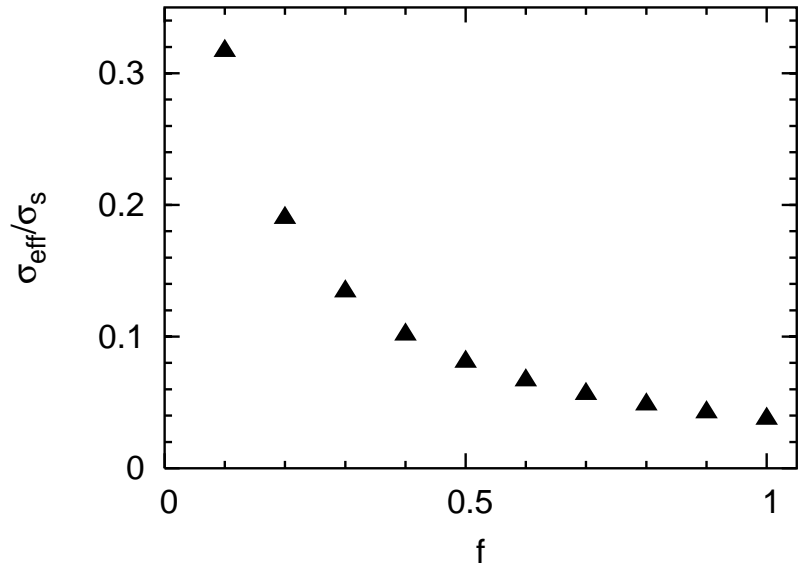}}\\
    \caption{$\sigma_{\mam{eff}}/\sigma_{\mam{s}} $ for $f=0.1, 0.2, \cdots, 0.9, 1.0$.
             The other parameters are the same as in Figure \ref{figure1}.
   }
    \label{figure9}
  \end{center}
 \end{figure}

\begin{figure}[htbp]
 \begin{center}
   \rotatebox{-90}{
     \includegraphics[width=8cm]{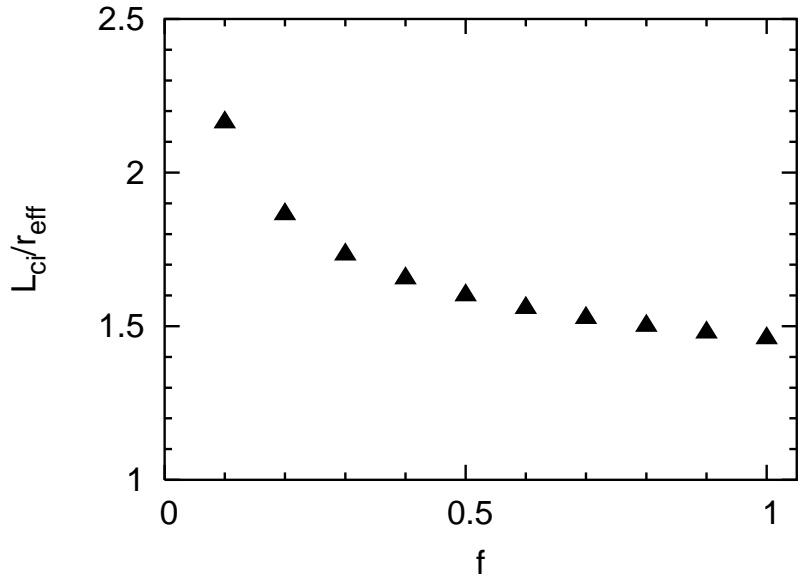}}\\
    \caption{$L_{\mam{ci}}/r_{\mam{eff}} $ for $f=0.1, 0.2, \cdots, 0.9, 1.0$.
             The other parameters are the same as in Figure \ref{figure1}.
   }
    \label{figure10}
  \end{center}
 \end{figure}

In Figure \ref{figure7} and Figure \ref{figure8},
we plot the absolute value of the negative charge distribution
$ f N/\sigma \rho_{\mam{m}}(z) $, the positive charge distribution $\rho_{\mam{ci}}(z)$,
the total charge distribution $\rho_{\mam{c}}(z)$, and the fitting GC solution (\ref{eq:r6}).
In Figure \ref{figure9} and \ref{figure10}, we plot $ \sigma_{\mam{eff}}/\sigma_{\mam{s}}, L_{\mam{ci}}/r_{\mam{eff}}$.

\begin{figure}[htbp]
 \begin{center}
   \rotatebox{-90}{
     \includegraphics[width=8cm]{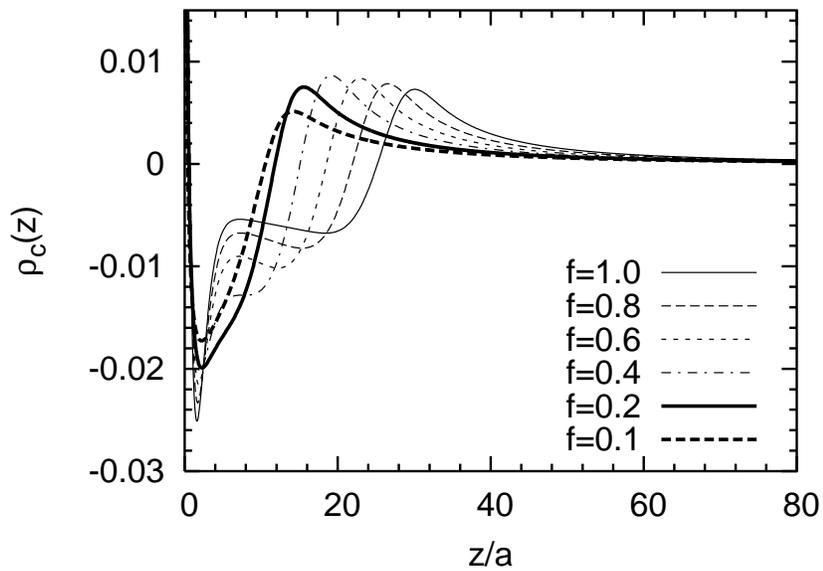}}\\
    \caption{The total charge distribution $\rho_{\mam{c}}(z)$ for 
             $f=0.1, 0.2, 0.4, 0.6, 0.8, 1.0$. The other parameters are the same as in Figure \ref{figure1}.}
    \label{figure6}
  \end{center}
 \end{figure}

For large values of $f \ge 0.5$, the ratio $\sigma_{\mam{eff}}/\sigma_{\mam{s}}$ is smaller 
than $ 0.10 $ and a large part of the negative charge of the brush is cancelled 
by the positive charge of counterions. This strong screening of the negative
charges of the brushes can be verified by comparing Figure \ref{figure6} and \ref{figure7}.
For decreasing $f$, the screening of the brush charge
becomes weaker and the counterions gradually escape from the brush.
This behavior of counterions is verified in Figure \ref{figure8}.

From Table \ref{table1}, we see that for $f\ge0.2$, the value of $\sigma_{\mam{eff}} \approx 0.12$ 
is almost constant.
As was already mentioned above, far away from the wall, the polyelectrolyte brush can be viewed as 
a charged plane located at $r_{\rm eff}$ with charge
density $-\sigma_{\mam{eff}} e$.
The charge density $-\sigma_{\mam{eff}} e$ is the `effective' charge of the brush which 
is not cancelled by the counterions.
Our numerical results show that there is an upper limit for this effective charge.

Next, in Figure \ref{figure6}, we plot the total charge distributions $\rho_{\mam{c}}(z)$ 
for various values of $f$.
Figure {\ref{figure6}} shows that the values of $\rho_{\mam{c}}(z)$
near the grafting plane are always positive for any $f$.
This is due to the depletion of the monomer density near the grafting plane.
The amount of positive charges of counterions is larger than that
of the negative charges of the brushes.
The minimum of $\rho_{\mam{c}}(z)$, which is negative, is located near the grafting
plane and an electric dipole layer is generated. 

The depth of the negative minimum of $\rho_{\mam{c}}(z)$ increases with $f$ and a
small peak in the absolute value of the positive charge $(fN/\sigma)
\rho_{\mam{m}}(z)$ appears for $f=1.0$  (see Figure {\ref{figure7}}).
As can be seen in  Figure {\ref{figure7}}, when $f$ increases,  in order to enforce locally charge neutrality, the monomer and counterion densities become closer. However, the depletion layer of the polymer near the wall is still present. 
This region is filled by the counterions, and this accumulation of positive charges is compensated by a peak in the polymer density.
Taniguchi \textit{et al.} \cite{taniguchi03}
obtained similar results. However the behavior of the
depth of their negative minimum is somewhat different from our results.

\section{Conclusion}

We formulated the self-consistent field theory for a planar polyelectrolyte brush
in a salt free solution in the presence of counterions.
We numerically solved the self-consistent field equations for the weak coupling case
and obtained the monomer density profile, the distribution of counterions,
and the total charge distribution.
We also studied the scaling relation for the brush height.  
For large values of the charge
fraction, our results correspond to those of the osmotic regime.  
We observe a weak dependence of the brush height on the grafting density.  
This is consistent with a recent study of the nonlinear osmotic regime.  
For small values of the charge 
fraction, our results correspond to those of the quasi-neutral regime.
We find a crossover phase between the osmotic regime and 
the quasi-neutral regime.

We studied the distribution of the total charge, and found that an
electric dipole layer appears near the grafting area.  We
studied the distribution of counterions, and fit it
by the Gouy-Chapman theory for a virtual charged wall at the location
$r_{\rm eff}$ with an effective charge density $\sigma_{\mam{eff}}$ .
We calculated the integrated amount of the charge of counterions
outside the brush and the thickness of the counterions outside the
brush by introducing an effective GC length ($l_{\mam{eff}}$).  For
large values of the charge fraction, the charges of the monomers are
strongly screened by the charge of the counterions, and the
counterions are trapped inside the brush.  For smaller values of the
charge fraction, the screening effects are weak and the counterions
extend far beyond the brush.  Far away from the grafted surface, the
polyelectrolyte brush can be viewed as a virtual charged plane with an
effective charge density $\sigma_{\mam{eff}}$.  We found that this
effective charge density has an upper limit and the rest of the
polymer charges are cancelled by the charges of the counterions.

\newpage

\begin{table}
 \begin{center}
 \caption{$2 h_{\mam{m}}, 2l_{\mam{ci}}, r_{\mam{eff}}, l_{\mam{eff}}, L_{\mam{ci}}, \sigma_{\mam{s}}
           , \sigma_{\mam{eff}}$ at different charge fractions $f$. The other parameters are the same 
as in Figure \ref{figure1}.}
 \label{table1}
 \begin{tabular}{|c|c|c|c|c|c|c|c|c|}
 \hline
 $f$ & $ 2h_{\mam{m}}/a $ & $ 2l_{\mam{ci}}/a $ & $ r_{\mam{eff}}/a $ & 
$ l_{\mam{eff}}/a$ & $ L_{\mam{ci}}/a $ & $\sigma_{\mam{s}}$ & $\sigma_{\mam{eff}}$\\
 \hline
 1.0  & 18.61  & 26.16   & 30.96 &  14.22   & 45.18  & 3.0  & 0.112  \\
 0.9  & 17.78  & 26.12   & 29.26 &  13.98   & 43.24  & 2.7  & 0.114  \\
 0.8  & 16.91  & 26.26   & 27.49 &  13.74   & 41.23  & 2.4  & 0.116  \\
 0.7  & 16.00  & 26.62   & 25.67 &  13.49   & 39.16  & 2.1  & 0.118   \\
 0.6  & 15.05  & 27.36   & 23.84 &  13.31   & 37.15  & 1.8  & 0.120   \\
 0.5  & 14.05  & 28.70   & 21.96 &  13.17   & 35.12  & 1.5  & 0.121   \\
 0.4  & 13.01  & 31.10   & 20.01 &  13.09   & 33.10  & 1.2  & 0.122  \\
 0.3  & 11.92  & 35.64   & 18.00 &  13.19   & 31.19  & 0.9  & 0.121   \\
 0.2  & 10.83  & 45.36   & 16.18 &  13.97   & 30.15  & 0.6  & 0.114   \\
 0.1  &  9.81  & 74.52   & 14.41 &  16.76   & 31.17  & 0.3  & 0.0950  \\
 \hline  
 \end{tabular}
 \end{center}
 \end{table}

\end{document}